\documentclass[prl,superscriptaddress,reprint]{revtex4-2}

\input{preamble.sty}

\begin{document}
\title{\titlestring}

\author{Giannis Thalassinos\hyperlink{corres}\textsuperscript{$^{\ast,\star}$}}
\affiliation{School of Science, RMIT University, Melbourne, VIC 3001, Australia}

\author{Alan G. Salek\hyperlink{corres}\textsuperscript{$^{\dagger,\star}$}}
\affiliation{School of Science, RMIT University, Melbourne, VIC 3001, Australia}
\affiliation{RMIT Microscopy and Microanalysis Facility, RMIT University, Melbourne, VIC 3001, Australia}

\author{Daniel Stavrevski\hyperlink{corres}\textsuperscript{$^{\star}$}}
\affiliation{School of Science, RMIT University, Melbourne, VIC 3001, Australia}

\author{Qiang Sun}
\affiliation{School of Science, RMIT University, Melbourne, VIC 3001, Australia}

\author{Mitchell O. de Vries}
\affiliation{School of Science, RMIT University, Melbourne, VIC 3001, Australia}
\affiliation{Quantum Machines Unit, Okinawa Institute of Science and Technology, Onna, Okinawa 904-0495, Japan}

\author{Colin M. MacRae}
\affiliation{CSIRO Mineral Resources, Microbeam Laboratory, Clayton, VIC 3168, Australia}

\author{Nicholas C. Wilson}
\affiliation{CSIRO Mineral Resources, Microbeam Laboratory, Clayton, VIC 3168, Australia}

\author{Andrew G. Tomkins}
\affiliation{School of Earth, Atmosphere and Environment, Monash University, Melbourne, VIC 3800, Australia}

\author{Dougal G. McCulloch\hyperlink{corres}\textsuperscript{$^{\ddagger}$}}
\affiliation{School of Science, RMIT University, Melbourne, VIC 3001, Australia}
\affiliation{RMIT Microscopy and Microanalysis Facility, RMIT University, Melbourne, VIC 3001, Australia}

\author{Andrew D. Greentree\hyperlink{corres}\textsuperscript{$^{\S}$}}
\affiliation{School of Science, RMIT University, Melbourne, VIC 3001, Australia}

\date{\today}

\begin{abstract} 
Lonsdaleite—hexagonal diamond—has only recently been proposed as a wide-bandgap host capable of supporting optically active point defects, but no such centres have yet been observed. 
Here we provide the first experimental evidence that lonsdaleite does in fact host photoluminescent color centres. 
In meteoritic lonsdaleite grains from the ureilite NWA7983, we identify a new defect, RU1, which exhibits bright and stable emission across $550–800$ nm, with optimal blue excitation ($\sim 455$~nm) and a peak at $\sim 700$~nm. 
Time-resolved photoluminescence reveals an excited-state lifetime of 14~ns with no detectable blinking, bleaching, or charge conversion. 
From the excitation–emission energetics we infer an unresolved zero-phonon line near 550~nm. 
Correlative electron microscopy confirms the lonsdaleite host lattice, and compositional analysis suggests N, Si, or Ni as plausible defect constituents. 
These results suggest lonsdaleite could become a new quantum-grade crystalline platform and indicate that hexagonal-diamond color centres may form a new and unexplored family of solid-state quantum emitters.
\end{abstract}

\maketitle

\section{Introduction}\label{sec1}

Crystals are typically transparent across spectral ranges defined by their bandgap. 
Color centers---imperfections arising from impurities or defects---introduce absorption and emission within the bandgap, producing the vivid colors of gemstones such as ruby, sapphire, and colored diamonds. 
Beyond their aesthetic value, color centers reveal defect formation, crystal symmetry, and act as the foundation of many emerging quantum technologies in crystals such as diamond, hexagonal boron nitride (hBN), and silicon carbide (SiC)~\cite{ACS+2011,scholten2024multispecies,castelletto2020silicon}.

Lonsdaleite, also known as hexagonal diamond, is a rare allotrope of carbon originally discovered in meteorite impact sites~\cite{FM67} and is also expected to have a wide bandgap~\cite{salehpour1990comparison}.
It has been synthesized using high-pressure high-temperature~\cite{BP67,CCG+2025}, high-pressure room-temperature~\cite{MWS+2020}, shock~\cite{HSK+2002}, and chemical vapor deposition~\cite{BBS+1995} processes. 
In all of these cases, lonsdaleite is present in small, sub-micron grains. 

Despite the availability of laboratory-grown samples, there is still much unknown about lonsdaleite. 
In particular, although it is predicted to be harder than diamond~\cite{PSZ+2009}, experiments suggest that the material has a similar indentation hardness to diamond~\cite{HST+2023}.  
Lonsdaleite is widely believed to be a hexagonal and pure sp$^3$ allotrope of carbon.
However, it has also been speculated that defect-free hexagonal diamond does not exist \textit{per se}, and that lonsdaleite is in fact a  disordered form of conventional, cubic diamond~\cite{NGA+2014,SMS2015}. 
Indeed, high-resolution electron microscopy studies have shown that, in many cases, lonsdaleite samples contain mixed cubic and hexagonal stacked diamond~\cite{Kulnitskiy2013, SALEK2025}. 
Regardless of its microstructure, like diamond, lonsdaleite is likely to have unique properties which, once characterized, could be exploited in a range of technologically important applications.

An interesting question is whether lonsdaleite might host color centers, analogous to those in diamond~\cite{zaitsev2013optical}. 
One of the most important color centers in diamond is the  nitrogen-vacancy (NV) color center~\cite{DMD+2013}, which has shown remarkable properties for quantum and bio applications~\cite{SCL+2014}. 
Given the similarities in structure between diamond and lonsdaleite, one might expect that lonsdaleite could host an NV center with similar properties to those in diamond. 
Indeed, modeling suggests that NV~\cite{SCX+2023,ACA2024,ACA2025,manian2025nitrogenvacancy} and other color centres~\cite{abdelghafar2026initio} may well exist.
Observing fluorescent color centers in lonsdaleite may help in understanding the symmetry groups present in lonsdaleite, thereby shedding light on its fundamental structure. 
In addition, these color centers may open up new applications in their own right. 

Here, we show cathodoluminescence (CL) and photoluminescence (PL) originating from a new color center, tentatively named RU1, originating from a meteoritic sample (NWA7983), a ureilite collected in Morocco~\cite{Ruzicka2015}.
This sample is known to contain both grains of lonsdaleite and diamond~\cite{SALEK2025}, which we characterize and distinguish using electron microscopy techniques. 
The RU1 centre shows broad PL with optimal excitation at around 455~nm, and peak emission at 700~nm.  
By assuming that the defect has a zero phonon line (ZPL) and that this is halfway between the peak for excitation and emission, we predict the (unresolved) ZPL to be around 550~nm.  
We observe bi-exponential photoluminescence decay with a long lifetime of 14~ns, and a fast lifetime of 2~ns and see no evidence photobleaching, blinking, or photo-induced charge state conversion in the samples.

\section{Methods}

A polished section of ureilite NWA 7983 was prepared using standard polishing methods~\cite{tomkins2022sequential,SALEK2025}, using diamond paste of decreasing particle size, from 6~\si{\um} down to 0.1~\si{\um}. 
Electron probe microanalysis (EPMA) was used to produce the elemental and CL maps of the ureilite, using a JEOL 8530F-CL HyperProbe at the CSIRO Mineral Resources Microbeam Laboratory in Melbourne. 
CL was collected using an optical grating spectrometer, xCLentV~\cite{Macrae2005} and a Bruker EDS with hardware and software integration of CL, EDS and other spectrometry developed by CSIRO~\cite{MacRae2013}. 
The optical spectrometer collected from \numrange{199}{972}~nm and had a 200~\si{\um} entrance slit. 
Operating conditions were 7~kV, 80~nA, and a dwell per pixel of 400~ms, with maps collected by scanning the stage with a step size and spot size of 500~nm.
The sample was cooled to 80~K using a GATAN-JEOL cold stage during mapping.

Photoluminescent properties of the ureilite sample were investigated using a custom-built confocal microscope with a tuneable laser (Fianium WhiteLase 400-SC, NKT Photonics, DEN) operating between \numrange{450}{520}~nm at 100~\si{\uW}, described elsewhere~\cite{thalassinos2025robust}. 
The sample was excited through a 100$\times$ objective (0.9 NA), with PL collected through the same objective and fiber-coupled.
An in-line 75:25 beam splitter sent the signal to a spectrometer (75~\%, SpectraPro SP-2500, Princeton Instruments) with a charge-coupled device (CCD) camera (PIXIS 100BR, Princeton Instruments) and to an avalanche photo-diode (APD, 25~\%, SPCM-AQRH-14-TR, Excelitas) synchronized with a correlator card (TimeHarp260, PicoQuant) for time-resolved measurements.

\section{Results and Discussion}
\subsection{Composition and cathodoluminescence}
\begin{figure}[t]
    \centering
    \includegraphics[width = 0.48\textwidth]{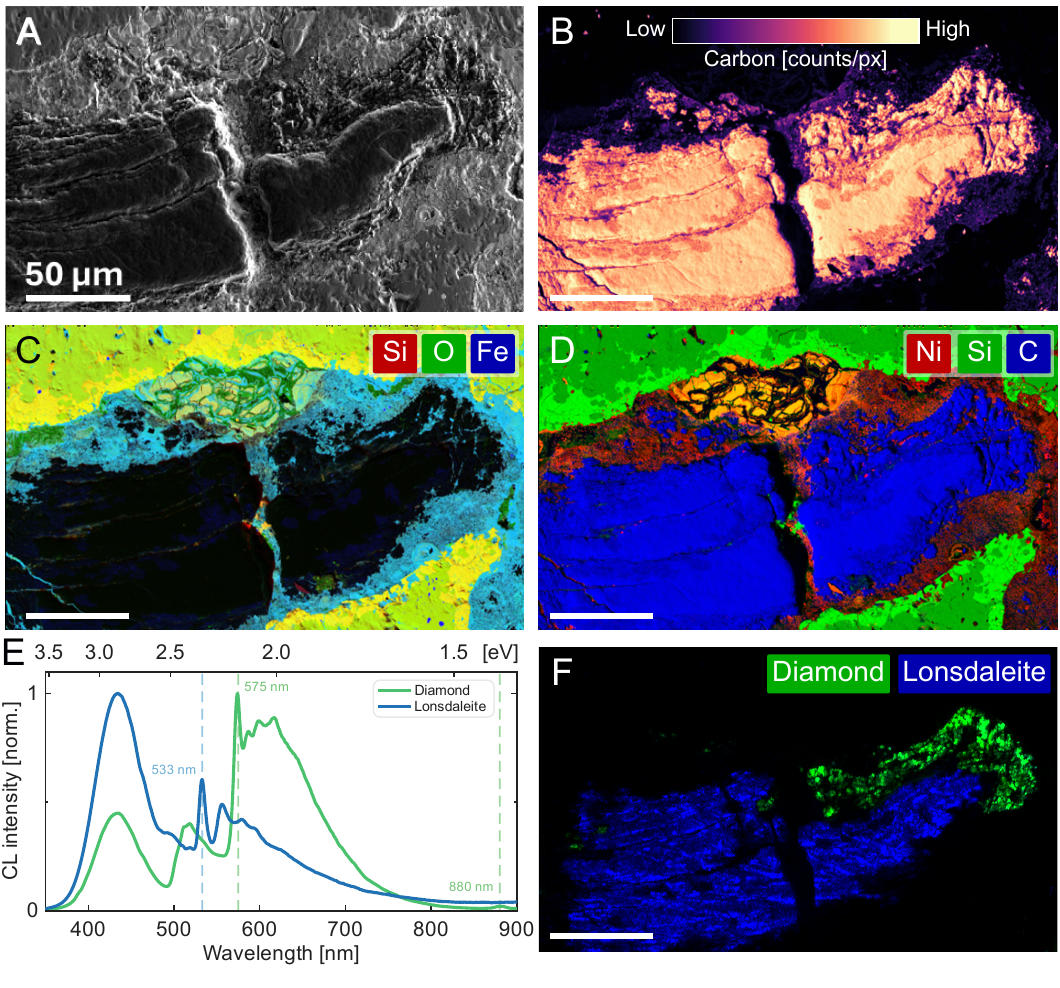}
    \caption{Electron microscopy analysis of a region of interest in ureilite NWA 7983. 
    \textbf{A:}~Scanning electron microscope image of a hard carbon-rich region. 
    \textbf{B:}~Elemental intensity map of carbon, showing that the smooth elongated grain and speckled crystals surrounding it, seen in (A), are pure carbon. 
    \textbf{C:}~Elemental map of the region of interest showing the surrounding minerals are comprised of Fe, O and Si. 
    \textbf{D:}~Elemental map highlighting the carbon-rich region, as well as surrounding Si and Ni. 
    \textbf{E:}~Cathodoluminescence spectra from lonsdaleite (black line) and diamond (red line) which have unique features. 
    The peak at 890~nm in the diamond spectrum likely arises from Ni-related color centers. 
    \textbf{F:}~De-convoluted CL response for diamond and lonsdaleite, with lonsdaleite projected in blue and diamond in green. 
    }
    \label{fig:1as}
\end{figure}

\Cref{fig:1as}A-D shows a secondary electron microscope image and elemental maps of a typical carbon-rich area with co-existing lonsdaleite, diamond, and graphite in the ureilite sample. 
The C-rich area is surrounded by olivine and pyroxene, as indicated by the high levels of Fe, Si and O (see our previous work for more details on this ureilite~\cite{tomkins2022sequential,SALEK2025}). 
The elemental maps also reveal high levels of Ni at the interface between the carbon grain and surrounding material. 
The process that leads to FeNi metal at this interface is a result of reduction of the surrounding olivine at high temperatures during decompression following the parent body’s catastrophic impact~\cite{Langendam2021,WG1993,SALEK2025}.  
\Cref{fig:1as}E shows typical CL spectra from lonsdaleite (blue line) and diamond (green line), which have strong CL emission with ZPLs at 533 and 575~nm, respectively. 
The emission at 575~nm, with accompanying phonon side-bands, is characteristic of the neutrally charged nitrogen vacancy (NV$^0$) centre in diamond~\cite{ACS+2011,thalassinos2025robust}. 

The diamond CL spectra has an additional feature at 880~nm, which can be attributed to a Ni  related color centre~\cite{zaitsev2013optical}. 
This is consistent with the presence of the surrounding Ni-rich material. 
Graphite does not produce a CL signal due to the lack of a band-gap. 
The characteristic CL peaks for lonsdaleite and diamond were used to produce the map shown in \cref{fig:1as}F. 
The diamond is concentrated around the edge of the predominantly lonsdaleite-rich carbon grain, with diamond grains sitting within graphite. 
Lonsdaleite is observed as elongated grains with morphology similar to preexisting graphite from which it formed~\cite{tomkins2022sequential,SALEK2025}.

 \begin{figure*}[tbh!]
    \centering
     \includegraphics[width = \textwidth]{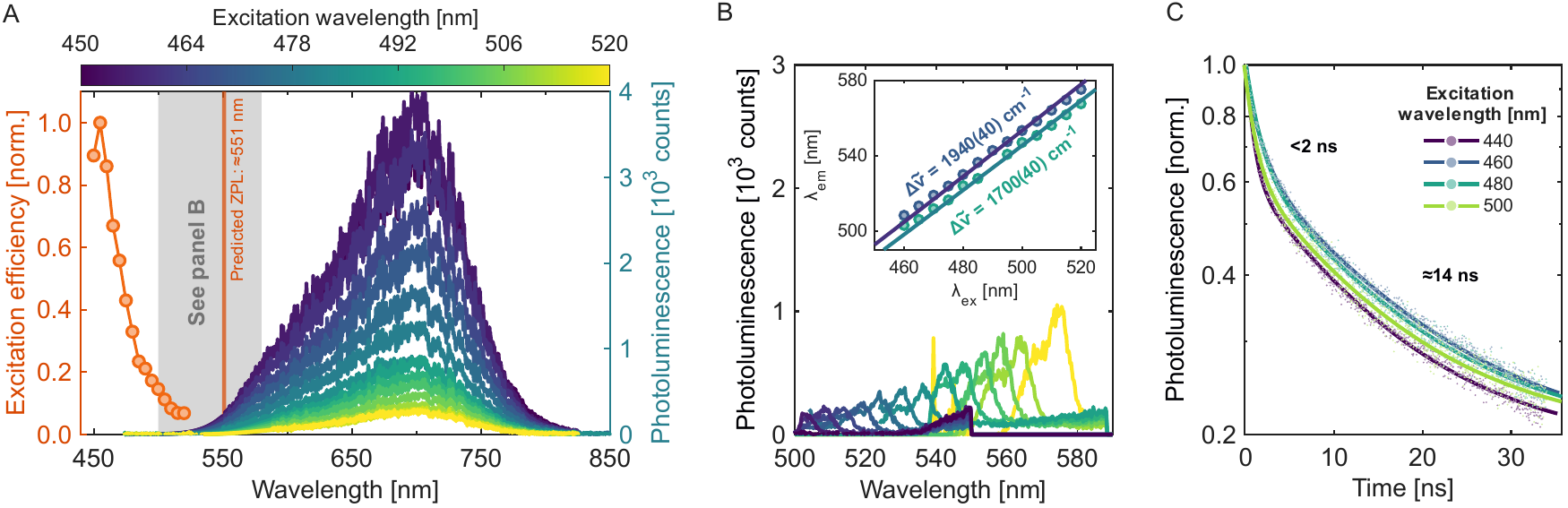}
    \caption{
    \textbf{A:}~Photoluminescence spectra of RU1 under excitation wavelengths from \numrange{450}{520}~nm (right axis) with accompanying excitation spectrum for the same wavelength range (left axis).
    The shaded region between \numrange{550}{600}~nm corresponds to the spectral window of the excitation-wavelength dependent signal, which was isolated into (B).
    \textbf{B:}~Isolated \lex-dependent signals extracted from the total emission spectrum in (A) by approximating the PL as the sum of to skewed Gaussians. 
    The inset shows the calculated Stokes shifts ($\Delta\nu$) based on the peak positions of two Gaussian fits.
    \textbf{C:}~Time resolved PL of RU1 for excitation wavelengths ranging \numrange{440}{500}~nm using a 20~MHz pulsed laser.}
    \label{fig:lambda_sweep}
\end{figure*}

\subsection{Photoluminescence characteristics}
We investigated the PL characteristics of multiple lonsdaleite grains under a scanning confocal microscope under excitation wavelengths $(\lambda_\text{ex}$) ranging \numrange{450}{520}~nm.
\Cref{fig:lambda_sweep} shows the PL properties of the region corresponding to lonsdaleite in \cref{fig:1as}, where we detected a new color center that we name RU1. 
RU1 exhibits broad emission (\cref{fig:lambda_sweep}A, right axis) between \numrange{550}{800}~nm, peaking at 700~nm, with no clear evidence of phononic structure, including no obvious ZPL.
As we sweep the excitation wavelength, we observe a change in total PL intensity, with no photochromic effects within this spectral window.

We estimate the excitation efficiency by fitting a double skewed Gaussian model to each PL spectrum, then integrating for total PL intensity for each excitation wavelength.
The normalized excitation efficiency of RU1 (\cref{fig:lambda_sweep}A, left axis), shows a peak excitation wavelength in the blue part of the spectrum at 455~nm.
Assuming that this centre has similar behavior to common defects in diamond, such as the nitrogen-vacancy centre, then we would expect the absorption and emission spectra to be mirrored around the zero phonon line.  
Under this assumption, we may interpret the spectra in \cref{fig:lambda_sweep}A as arising due to a zero phonon line halfway between the peaks of absorption and emission, with the large features corresponding to the phononic branching ratio.  
We note that it is not uncommon for the phonon sidebands to be unresolved in nano- and micro-crystalline samples.
With a peak excitation wavelength of 455~nm (2.725~eV) and maximum emission at 700~nm (1.771~eV), we expect to find the ZPL at 2.248~eV ($\approx$550~nm).
While not an exact match, this is consistent with some predictions of the ZPL of the NV$^0$ center in lonsdaleite~\cite{ACA2025,manian2025nitrogenvacancy}. 

Additionally, we detect a weaker, $\lambda_\text{ex}$-dependent, signal within the spectral window ranging \numrange{500}{580}~nm (shaded region in \cref{fig:lambda_sweep}A) which we have separated into \cref{fig:lambda_sweep}B.
To separate the signals, we used the double-Skewed Gaussian fit to subtract the outliers within this spectral window to return a `pure' PL spectrum, and vice-versa to obtain the signal in \cref{fig:lambda_sweep}B.
The original non-separated spectra can be found in SI fig. S1.

We found that the peak emission wavelength ($\lambda_\text{em}$) shifts proportionally to the excitation wavelength.
By applying a double Gaussian fit to each spectrum, we calculated the peak emission wavelengths for the pair of peaks under each excitation wavelength.
From this, we quantify the Stokes shift ($\Delta\tilde{\nu}$) using the expression
~
\begin{equation}\label{eq:raman}
    \Delta\tilde{\nu} = \left( \frac{1}{\lambda_\text{ex}}  - \frac{1}{\lambda_\text{em}} \right),
\end{equation}
which gives us Stokes shifts of $\Delta\tilde{\nu}$~=~\num{1700(40)} and \num{1940(40)}~\si{\per\cm} (see the inset of \cref{fig:lambda_sweep}B).
These features are not typically associated with lonsdaleite, which exhibits Raman shifts in the range of $\approx$\numrange{1200}{1500}~\si{\per\cm}~\cite{smith2009uv,SALEK2025,yang2025synthesis}.
However, previous work shows that the peaks at 1700 and 1940~\si{\per\cm} also appear alongside the normal lonsdaleite Raman shifts in these ureilite samples~\cite{SALEK2025}.

 \begin{figure*}[tb]
    \centering
    \includegraphics[width=1\textwidth]{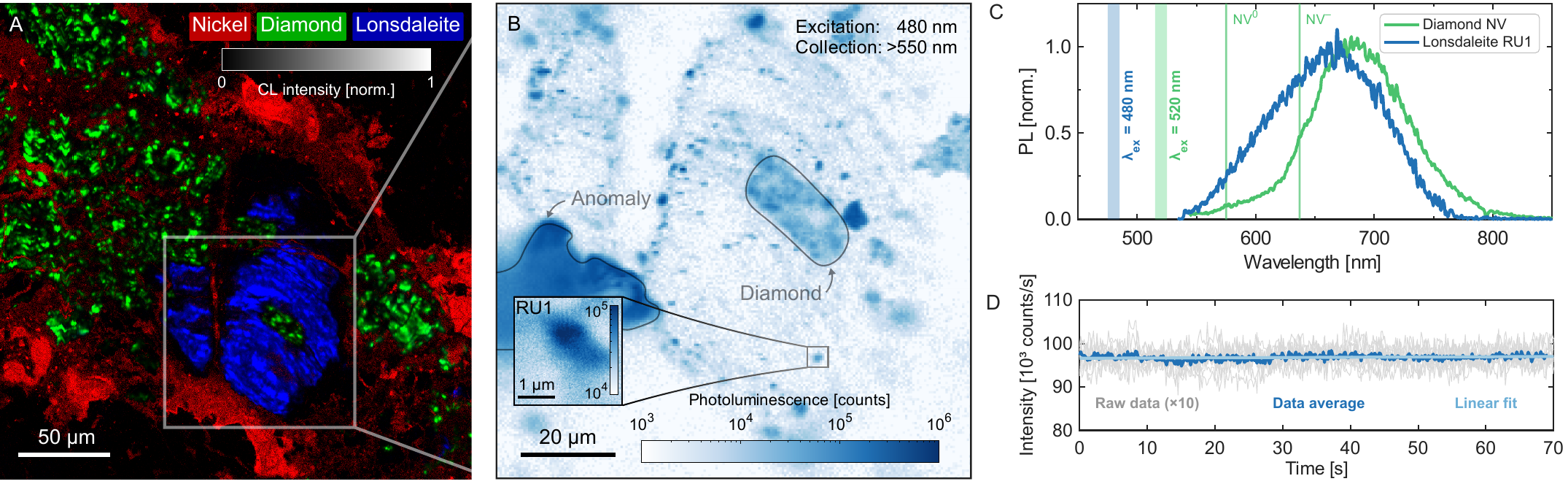}
    \caption{
    \textbf{A:}~Cathodoluminescence map of a second region containing diamond (green), lonsdaleite (blue), and nickel (red). 
    The CL spectra (\cref{fig:1as}E) was de-convoluted to extract each component.
    The boxed region marked along the bottom half of the map corresponds to the region shown in (B). 
    \textbf{B:}~PL confocal map of the sub-region marked in (A), using 480~nm excitation with collection above 550~nm.
    We note that the bright feature in the center-left of the map is an anomaly. 
    The inset shows a zoomed in map of the RU1 color center. 
    \textbf{C:}~PL spectra of lonsdaleite RU1 compared to PL from nearby diamond which can be attributed to NV centers. 
    Lonsdaleite and diamond were excited with 480 and 520~nm, respectively, marked by the shaded regions. 
    Solid vertical lines indicate the locations of the diamond NV zero-phonon lines.
    \textbf{D:}~PL intensity of lonsdaleite RU1 under continuous pulsed excitation with average power of 500~\si{\uW}, averaged (dark blue) over 10 sequential measurements (gray). 
    A linear fit (light blue) suggests that RU1 is highly photostable. 
    }
    \label{fig:lons_v_diamond}
\end{figure*}

We used a 20~MHz pulsed excitation source to probe the time-resolved PL characteristics of RU1 for excitation wavelengths 440, 460, 480, and 500~nm. 
\Cref{fig:lambda_sweep}C shows the time-resolved PL decay of RU1 across each $\lambda_\text{ex}$, normalized to maximum intensity.
Here, we observe a bi-exponential decay with a fast ($<$2~ns) and slow component $\approx$14~ns with roughly \numrange{50}{60}~\% of the signal originating from the slow component. 
The lifetime remains unchanged across the different excitation wavelengths. 

We further investigated a second lonsdaleite sample within the same meteoritic specimen. 
\Cref{fig:lons_v_diamond}A shows a CL map approximately 250 $\times$ 250~\si{\um\squared}, which was generated using the same CL spectral features shown in \cref{fig:1as}E. 
Here, we find a lonsdaleite grain which spans $\approx$100~\si{\um}, which is surrounded by diamonds that are generally much brighter under CL. 
Detailed elemental composition maps can be found in SI fig. S2.
Using 480~nm excitation, we investigated the lonsdaleite region marked by the white box in \cref{fig:lons_v_diamond}A under our confocal microscope to obtain a PL map, shown in \cref{fig:lons_v_diamond}B. 
Here, we observe fluorescent features across the grain, many appearing along the edges of the sample.
We note that not all PL originating from this region corresponds to RU1, and we highlight one particularly bright region (center left) which is likely an anomaly and cannot be attributed to either lonsdaleite or diamond. 
Regardless, we identify RU1 in this area of interest, shown in the inset in \cref{fig:lons_v_diamond}B.

The PL spectrum of RU1 can be seen again in \cref{fig:lons_v_diamond}C, which largely matches what was observed in the previous sample as shown in \cref{fig:lambda_sweep}.
We compare this signal to PL originating from nearby diamond, which shows PL characteristic of the NV center in diamond under 532~nm excitation~\cite{DMD+2013}. 
Notably, the NV center also lacks distinguishable ZPLs at 575 and 637~nm, which is common for NV centers in nanoparticles and/or near the diamond surface~\cite{BGN+2015}.
While both signals show significant overlap, the RU1 emission spectrum cannot be reconstructed as a combination of pure diamond NV$^0$ and NV$^-$ spectra~\cite{thalassinos2025robust}, suggesting that RU1 is an entirely different and new color center.

Finally, RU1 appears to be highly photostable, as shown in \cref{fig:lons_v_diamond}D, which shows an averaged time trace (dark blue) of PL collected from RU1 under constant excitation using 480~nm at 500~\si{\uW}, averaged over 10 sequential measurements (gray). 
Across all measurements, we observed an average count rate of \num{96.8(6)e3}~counts/s.
A linear fit shows a statistically insignificant rate of change in the count rate of \num{6(3)}~counts/s$^2$, demonstrating that RU1 is highly photostable over this time span, with no evidence of photobleaching or blinking.

\section{Discussion}
The composition of the RU1 color center is unknown, yet there may be a few candidates based on the available information. 
We detect O, Si, Ni, and Fe, with Ni in particular appearing to surround the lonsdaleite as per \cref{fig:1as}D. 
Both Si and Ni are known to form color centers in diamond~\cite{zaitsev2013optical}, such as the silicon-vacancy (SiV) and positively charged substitutional nickle (Ni$_\text{s}^+$) color centers~\cite{shames2020nearinfrared}, which generally have weak electron-phonon coupling, leading to narrow emission spectra. 

We also detect PL characteristic of NV centers in diamond (D-NV) in the diamonds surrounding the lonsdaleite, indicating the presence of nitrogen in the region. 
Modeling suggests that lonsdaleite could host a nitrogen vacancy center (L-NV)~\cite{SCX+2023,ACA2024,ACA2025,manian2025nitrogenvacancy} with emission in the visible and near-infrared (NIR) spectrum, depending on charge state.

Similar to D-NV, RU1 has a broad emission spectrum spanning $\approx$\numrange{550}{800}~nm, peaking at around 700~nm.
However, unlike D-NV which shows overlapping emission from its two charge states, D-NV$^0$ and D-NV$^-$, we have not observed photochromism from RU1 in the investigated spectral region, showing only a change in total PL intensity as a function of excitation wavelength. 
If RU1 is the theoretically predicted L-NV, we may expect to see the negative charge state (L-NV$^-$) further in the near-infrared part of the spectrum, as we see in diamond's N$_2$V defect~\cite{johnson2025nitrogenvacancynitrogen}. 

We note that the CL emission spectrum observed in lonsdaleite shows a ZPL at 533~nm with a phonon side-band that is blue-shifted compared to both D-NV$^0$ and the observed PL emission of RU1. 
In diamond, CL spectroscopy leads to an electron induced charge carrier mechanism which converts D-NV$^-$ into D-NV$^0$~\cite{sola-garcia2020electroninduced}.
This effect can be seen between CL and PL spectra of the diamond samples, which show either exclusively D-NV$^0$ (CL) or a mix of both charge states (PL).
Assuming that the same mechanism occurs in lonsdaleite, one potential explanation for the differences between the CL and PL spectra of lonsdaleite may be related to different charge states of RU1. 

Compared to many other color centers, RU1 exhibits a long PL lifetime on the order of 14~ns, similar to D-NV which ranges between $\approx$\numrange{12}{24}~ns depending on the NV charge state and the local refractive index~\cite{DMD+2013,inam2013emission}.
We also observe a fast-decaying component on the order of 2~ns. 
A similar behavior can be observed in D-NV, which is usually attributed to increases in non-radiative decay pathways~\cite{reineck2019not}. 
The lack of distinguishable ZPLs in either RU1 or D-NV in our meteorite sample may allude to poor crystalline qualities, consistent with previous transmission electron microscope observations~\cite{SALEK2025}, which may contribute to the fast-decaying component observed in RU1. 

Finally, RU1 demonstrates high photostability, which is an important property of D-NV~\cite{DMD+2013} and other color centers~\cite{johnson2025nitrogenvacancynitrogen,shames2020nearinfrared}, which contributes to their implementation in emerging quantum technologies~\cite{degen_quantum_2017}. 
If RU1 is indeed the theoretically predicted L-NV, it may open up new avenues in quantum sensing, with potentially improved capabilities in vector magnetometry compared to D-NV due to the different bond lengths  of L-NV center along different directions~\cite{manian2025nitrogenvacancy}. 
However, more research is required to determine the composition, structure, and sensing capabilities of RU1. 

\section{Conclusion}\label{sec13}
As a wide band-gap semiconductor, lonsdaleite might be expected to host optically-active color centers, similar to diamond.
For the first time, we demonstrate that lonsdaleite does indeed host fluorescent color centers. 
The RU1 color center shows stable emission in the visible to near-infrared spectrum peaking at 700~nm with a long excited state lifetime of $\approx$14~ns.
Based on compositional analysis of the surrounding material, and some supporting evidence from diamonds present, possible candidates for the composition of this lonsdaleite center may include N, Si or Ni. 

With the development of new fabrication methodologies for the creation of lonsdaleite, the existence of color centres, especially if they are shown to have similar properties to nitrogen-vacancy in diamond, means that lonsdaleite may become a new solid state quantum material.

\section{Acknowledgments}
The authors gratefully acknowledge funding from the Australian Research Council (ARC) through LEIF Grant No. LE130100087.
G.T. acknowledges support through the ARC Linkage Grant No. LP210300230 in collaboration with Diamond Defence Pty Ltd.

\section*{Author Information}
\hypertarget{corres}{}
\noindent\textsuperscript{$\star$}{\small{These authors contributed equally to this work.}}

\noindent
\textbf{Correspondence:} \\
{\small
\textsuperscript{$\ast$}Giannis Thalassinos --- \href{mailto:giannis.thalassinos@rmit.edu.au}{giannis.thalassinos@rmit.edu.au} \\
\textsuperscript{$\dagger$}Alan G. Salek --- \href{mailto:alan.salek@rmit.edu.au}{alan.salek@rmit.edu.au} \\
\textsuperscript{$\ddagger$}Dougal  G. Mcculloch --- \href{mailto:dougal.mcCulloch@rmit.edu.au}{dougal.mcCulloch@rmit.edu.au} \\
\textsuperscript{$\S$}Andrew D. Greentree --- \href{mailto:andrew.greentree@rmit.edu.au}{andrew.greentree@rmit.edu.au}
}

\section*{References}
\bibliography{Lonsdaleite}

@article{FM67,
	author = {Frondel,Clifford and Marvin, Ursula B.},
	date = {1967/05/01},
	doi = {10.1038/214587a0},
	id = {FRONDEL1967},
	isbn = {1476-4687},
	journal = {Nature},
	number = {5088},
	pages = {587--589},
	title = {Lonsdaleite, a Hexagonal Polymorph of Diamond},
	volume = {214},
	year = {1967}}

@article{BP67,
    author = {Bundy, F. P. and Kasper, J. S.},
    title = "{Hexagonal Diamond—A New Form of Carbon}",
    journal = {The Journal of Chemical Physics},
    volume = {46},
    number = {9},
    pages = {3437-3446},
    year = {1967},
    month = {05},
    issn = {0021-9606},
    doi = {10.1063/1.1841236},
}

@article{HSK+2002,
    author = {He, Hongliang and Sekine, T. and Kobayashi, T.},
    title = "{Direct transformation of cubic diamond to hexagonal diamond}",
    journal = {Applied Physics Letters},
    volume = {81},
    number = {4},
    pages = {610-612},
    year = {2002},
    month = {07},
    issn = {0003-6951},
    doi = {10.1063/1.1495078},
}

@article{MWS+2020,
author = {McCulloch, Dougal G. and Wong, Sherman and Shiell, Thomas B. and Haberl, Bianca and Cook, Brenton A. and Huang, Xingshuo and Boehler, Reinhard and McKenzie, David R. and Bradby, Jodie E.},
title = {Investigation of Room Temperature Formation of the Ultra-Hard Nanocarbons Diamond and Lonsdaleite},
journal = {Small},
volume = {16},
number = {50},
pages = {2004695},
doi = {https://doi.org/10.1002/smll.202004695},
year = {2020}
}

@article{BBS+1995,
    author = {Bhargava, Sanjay and Bist, H. D. and Sahli, S. and Aslam, M. and Tripathi, H. B.},
    title = "{Diamond polytypes in the chemical vapor deposited diamond films}",
    journal = {Applied Physics Letters},
    volume = {67},
    number = {12},
    pages = {1706-1708},
    year = {1995},
    month = {09},
    issn = {0003-6951},
    doi = {10.1063/1.115023},
}

@article{PSZ+2009,
  title = {Harder than Diamond: Superior Indentation Strength of Wurtzite BN and Lonsdaleite},
  author = {Pan, Zicheng and Sun, Hong and Zhang, Yi and Chen, Changfeng},
  journal = {Phys. Rev. Lett.},
  volume = {102},
  issue = {5},
  pages = {055503},
  numpages = {4},
  year = {2009},
  month = {Feb},
  publisher = {American Physical Society},
  doi = {10.1103/PhysRevLett.102.055503},
}

@article{NGA+2014,
	author = {N{\'e}meth, P{\'e}ter and Garvie, Laurence A. J. and Aoki, Toshihiro and Dubrovinskaia, Natalia and Dubrovinsky, Leonid and Buseck, Peter R.},
	date = {2014/11/20},
	doi = {10.1038/ncomms6447},
	id = {N{\'e}meth2014},
	isbn = {2041-1723},
	journal = {Nature Communications},
	number = {1},
	pages = {5447},
	title = {Lonsdaleite is faulted and twinned cubic diamond and does not exist as a discrete material},
	volume = {5},
	year = {2014}}

@article{SMS2015,
title = {Extent of stacking disorder in diamond},
journal = {Diamond and Related Materials},
volume = {59},
pages = {69-72},
year = {2015},
issn = {0925-9635},
doi = {https://doi.org/10.1016/j.diamond.2015.09.007},
author = {Christoph G. Salzmann and Benjamin J. Murray and Jacob J. Shephard},
}

@article{HST+2023,
    author = {Huang, Xingshuo and Salek, Alan and Tomkins, Andrew G. and MacRae, Colin M. and Wilson, Nicholas C. and McCulloch, Dougal G. and Bradby, Jodie E.},
    title = "{Hardness of nano- and microcrystalline lonsdaleite}",
    journal = {Applied Physics Letters},
    volume = {122},
    number = {8},
    pages = {081902},
    year = {2023},
    month = {02},
    issn = {0003-6951},
    doi = {10.1063/5.0138911},
}

@article{DMD+2013,
title = {The nitrogen-vacancy colour centre in diamond},
journal = {Physics Reports},
volume = {528},
number = {1},
pages = {1-45},
year = {2013},
note = {The nitrogen-vacancy colour centre in diamond},
issn = {0370-1573},
doi = {https://doi.org/10.1016/j.physrep.2013.02.001},
author = {Marcus W. Doherty and Neil B. Manson and Paul Delaney and Fedor Jelezko and Jörg Wrachtrup and Lloyd C.L. Hollenberg},
}

@article{SCL+2014,
author = {Schirhagl, Romana and Chang, Kevin and Loretz, Michael and Degen, Christian L.},
title = {Nitrogen-Vacancy Centers in Diamond: Nanoscale Sensors for Physics and Biology},
journal = {Annual Review of Physical Chemistry},
volume = {65},
number = {1},
pages = {83-105},
year = {2014},
doi = {10.1146/annurev-physchem-040513-103659},
}

@article{SCX+2023,
title = {Fluorine and oxygen terminated hexagonal diamond (100) surfaces for nitrogen-vacancy based quantum sensors},
journal = {Diamond and Related Materials},
volume = {137},
pages = {110064},
year = {2023},
issn = {0925-9635},
doi = {https://doi.org/10.1016/j.diamond.2023.110064},
author = {Zhaolong Sun and Bo Cui and Wencui Xiu and Pingping Liang and Qimeng Liu and Nan Gao and Hongdong Li},
}

@book{zaitsev2013optical,
  title={Optical properties of diamond: a data handbook},
  author={Zaitsev, Alexander M},
  publisher={Springer Science \& Business Media},
address = {Berlin, Heidelberg},
doi = {10.1007/978-3-662-04548-0},
isbn = {978-3-642-08585-7},
publisher = {Springer Berlin Heidelberg},
year = {2001}
}

@article{Ruzicka2015,
author = {Ruzicka, Alex and Grossman, Jeffrey and Bouvier, Audrey and Herd, Christopher D. K. and Agee, Carl B.},
doi = {10.1111/maps.12491},
issn = {1086-9379},
journal = {Meteoritics \& Planetary Science},
month = {sep},
number = {9},
pages = {1662--1662},
title = {{The Meteoritical Bulletin, No. 102}},
volume = {50},
year = {2015}
}

@article{tomkins2022sequential,
  title={Sequential lonsdaleite to diamond formation in ureilite meteorites via in situ chemical fluid/vapor deposition},
  author={Tomkins, Andrew G and Wilson, Nicholas C and MacRae, Colin and Salek, Alan and Field, Matthew R and Brand, Helen EA and Langendam, Andrew D and Stephen, Natasha R and Torpy, Aaron and Pint{\'e}r, Zsanett and others},
  journal={Proceedings of the National Academy of Sciences},
  volume={119},
  number={38},
  pages={e2208814119},
  year={2022},
  publisher={National Acad Sciences},
  doi = {10.1073/pnas.2208814119}
}

@Article{ACA2024,
  author    = {Abdelghafar, Khaled A. and Choi, Daniel S and Askar, Khalid},
  journal   = {Journal of Physics D: Applied Physics},
  title     = {A density functional theory study of nitrogen vacancy center in lonsdaleite},
  year      = {2024},
  issn      = {1361-6463},
  pages     = {025113},
  volume    = {58},
  number    = {2},
  doi       = {10.1088/1361-6463/ad85ef},
  publisher = {IOP Publishing},
}

@Article{ACA2025,
  author    = {Abdelghafar, Khaled A. and Choi, Daniel S. and Askar, Khalid A.},
  journal   = {Scientific Reports},
  title     = {Luminescence lineshapes of nitrogen vacancy center in lonsdaleite and dual structure of diamond/lonsdaleite: a DFT study},
  year      = {2025},
  issn      = {2045-2322},
  number    = {1},
  volume    = {15},
  pages     = {15334},
  doi       = {10.1038/s41598-025-96242-w},
  publisher = {Springer Science and Business Media LLC},
}

@article{Kulnitskiy2013,
  title = {Polytypes and twins in the diamond–lonsdaleite system formed by high-pressure and high-temperature treatment of graphite},
  volume = {69},
  ISSN = {2052-5206},
  DOI = {10.1107/s2052519213021234},
  number = {5},
  journal = {Acta Crystallographica Section B Structural Science Crystal Engineering and Materials},
  publisher = {International Union of Crystallography (IUCr)},
  author = {Kulnitskiy,  Boris and Perezhogin,  Igor and Dubitsky,  Gennady and Blank,  Vladimir},
  year = {2013},
  month = sep,
  pages = {474–479}
}

@article{Macrae2005,
  author    = {Macrae, Colin M. and Wilson, Nicholas C. and Johnson, Sally A. and Phillips, Peter L. and Otsuki, Masayuki},
  title     = {{Hyperspectral mapping--combining cathodoluminescence and X-ray collection in an electron microprobe}},
  journal   = {Microscopy Research and Technique},
  volume    = {67},
  number    = {5},
  pages     = {271--277},
  year      = {2005},
  month     = {aug},
  doi       = {10.1002/jemt.20205},
  issn      = {1059-910X},
  publisher = {Wiley-Liss},
  address   = {United States}
}

@article{Langendam2021,
author = {Langendam, Andrew D. and Tomkins, Andrew G. and Evans, Katy A. and Wilson, Nicholas C. and MacRae, Colin M. and Stephen, Natasha R. and Torpy, Aaron},
title = {CHOS gas/fluid-induced reduction in ureilites},
journal = {Meteoritics \& Planetary Science},
volume = {56},
number = {11},
pages = {2062-2082},
doi = {https://doi.org/10.1111/maps.13755},
year = {2021}
}

@article{SALEK2025,
title = {Investigation of carbon phases in ureilite meteorites using electron microscopy techniques},
journal = {Carbon Trends},
volume = {20},
pages = {100534},
year = {2025},
issn = {2667-0569},
doi = {https://doi.org/10.1016/j.cartre.2025.100534},
url = {https://www.sciencedirect.com/science/article/pii/S2667056925000847},
author = {Alan G. Salek and Andrew G. Tomkins and Nicholas C. Wilson and Colin M. MacRae and Brock M. Nicholas and Dougal G. McCulloch},
}

@article{WG1993,
author = {Walker, David and Grove, Tim},
title = {Ureilite smelting},
journal = {Meteoritics},
volume = {28},
number = {5},
pages = {629-636},
doi = {https://doi.org/10.1111/j.1945-5100.1993.tb00633.x},
year = {1993}
}

@article{MacRae2013,
  author    = {MacRae, C. M. and Wilson, N. C. and Torpy, A.},
  title     = {{Hyperspectral cathodoluminescence}},
  journal   = {Mineralogy and Petrology},
  volume    = {107},
  number    = {3},
  pages     = {429--440},
  year      = {2013},
  month     = {jun},
  doi       = {10.1007/s00710-013-0272-8},
  issn      = {1438-1168},
}

@article{reineck2019not,
  title = {Not {{All Fluorescent Nanodiamonds Are Created Equal}}: {{A Comparative Study}}},
  shorttitle = {Not {{All Fluorescent Nanodiamonds Are Created Equal}}},
  author = {Reineck, Philipp and Trindade, Leevan Fremiot and Havlik, Jan and Stursa, Jan and Heffernan, Ashleigh and Elbourne, Aaron and Orth, Antony and Capelli, Marco and Cigler, Petr and Simpson, David A. and Gibson, Brant C.},
  year = {2019},
  journal = {Particle \& Particle Systems Characterization},
  volume = {36},
  number = {3},
  pages = {1900009},
  issn = {1521-4117},
  doi = {10.1002/ppsc.201900009},
  urldate = {2019-11-11},
  copyright = {{\copyright} 2019 WILEY-VCH Verlag GmbH \& Co. KGaA, Weinheim},
  langid = {english}
}

@article{inam2013emission,
	author = {Inam, Faraz A. and Grogan, Michael D. W. and Rollings, Mathew and Gaebel, Torsten and Say, Jana M. and Bradac, Carlo and Birks, Tim A. and Wadsworth, William J. and Castelletto, Stefania and Rabeau, James R. and Steel, Michael. J.},
	date = {2013/05/28},
	doi = {10.1021/nn304202g},
	isbn = {1936-0851},
	journal = {ACS Nano},
	journal1 = {ACS Nano},
	journal2 = {ACS Nano},
	month = {05},
	number = {5},
	pages = {3833--3843},
	publisher = {American Chemical Society},
	title = {Emission and Nonradiative Decay of Nanodiamond NV Centers in a Low Refractive Index Environment},
	type = {doi: 10.1021/nn304202g},
	volume = {7},
	year = {2013},
	year1 = {2013}}

@article{shames2020nearinfrared,
  title = {Near-{{Infrared Fluorescence}} from {{Silicon-}} and {{Nickel-Based Color Centers}} in {{High-Pressure High-Temperature Diamond Micro-}} and {{Nanoparticles}}},
  author = {Shames, Alexander I. and Dalis, Adamos and Greentree, Andrew D. and Gibson, Brant C. and Abe, Hiroshi and Ohshima, Takeshi and Shenderova, Olga and Zaitsev, Alexander and Reineck, Philipp},
  year = {2020},
  journal = {Advanced Optical Materials},
  volume = {8},
  number = {23},
  pages = {2001047},
  issn = {2195-1071},
  doi = {10.1002/adom.202001047},
  langid = {english}
}

@article{smith2009uv,
  title = {{{UV}} and {{VIS Raman}} Spectra of Natural Lonsdaleites: {{Towards}} a Recognised Standard},
  shorttitle = {{{UV}} and {{VIS Raman}} Spectra of Natural Lonsdaleites},
  author = {Smith, David C. and Godard, Gaston},
  year = {2009},
  journal = {Spectrochimica Acta Part A: Molecular and Biomolecular Spectroscopy},
  shortjournal = {Spectrochimica Acta Part A: Molecular and Biomolecular Spectroscopy},
  volume = {73},
  number = {3},
  pages = {428--435},
  issn = {1386-1425},
  doi = {10.1016/j.saa.2008.10.025}
}

@article{thalassinos2025robust,
  title = {Robust Quantification of the Diamond Nitrogen-Vacancy Center Charge State via Photoluminescence Spectroscopy},
  author = {Thalassinos, G. and McCloskey, D. J. and Mameli, A. and Healey, A. J. and Pattinson, C. and Simpson, D. and Gibson, B. C. and Stacey, A. and Dontschuk, N. and Reineck, P.},
  date = {2025-10-01},
  journal = {APL Photonics},
  volume = {10},
  number = {10},
  pages = {101102},
  issn = {2378-0967},
  doi = {10.1063/5.0284237},
  year = {2025},
}

@article{yang2025synthesis,
  title = {Synthesis of Bulk Hexagonal Diamond},
  author = {Yang, Liuxiang and Lau, Kah Chun and Zeng, Zhidan and Zhang, Dongzhou and Tang, Hu and Yan, Bingmin and Niu, Guoliang and Gou, Huiyang and Yang, Yanping and Yang, Wenge and Luo, Duan and Mao, Ho-kwang},
  year = {2025},
  journal = {Nature},
  volume = {644},
  pages = {370--375},
  publisher = {Nature Publishing Group},
  issn = {1476-4687},
  doi = {10.1038/s41586-025-09343-x},
  langid = {english},
}

@article{manian2025nitrogenvacancy,
    title = {Nitrogen-vacancy centre in lonsdaleite: a novel nanoscale sensor?},
    doi = {10.1039/D5CP01856K},
    journal = {Physical Chemistry Chemical Physics},
    author = {Manian, Anjay and Vries, Mitchell Owen de and Stavrevski, Daniel and Sun, Qiang and P Russo, Salvy and Greentree, Andrew},
    year = {2025},
    pages = {},
    volume = {}
}

@article{johnson2025nitrogenvacancynitrogen,
	author = {Johnson, Brett C. and de Vries, Mitchell O. and Healey, Alexander J and Capelli, Marco and Manian, Anjay and Thalassinos, Giannis and Abraham, Amanda N. and Hapuarachchi, Harini and Luo, Tingpeng and Mochalin, Vadym N. and Jeske, Jan and Cole, Jared H. and Russo, Salvy and Gibson, Brant C. and Stacey, Alastair and Reineck, Philipp},
	date = {2025/05/27},
	doi = {10.1021/acsnano.4c18283},
	isbn = {1936-0851},
	journal = {ACS Nano},
	month = {05},
	number = {20},
	pages = {19046--19056},
	publisher = {American Chemical Society},
	title = {The Nitrogen-Vacancy-Nitrogen Color Center: A Ubiquitous Visible and Near-Infrared-II Quantum Emitter in Nitrogen-Doped Diamond},
	type = {doi: 10.1021/acsnano.4c18283},
	url = {https://doi.org/10.1021/acsnano.4c18283},
	volume = {19},
	year = {2025},
	year1 = {2025}}

@article{degen_quantum_2017,
    title = {Quantum sensing},
    volume = {89},
    doi = {10.1103/RevModPhys.89.035002},
    number = {3},
    urldate = {2024-03-27},
    journal = {Reviews of Modern Physics},
    author = {Degen, C. L. and Reinhard, F. and Cappellaro, P.},
    year = {2017},
    pages = {035002},
}

@article{ACS+2011,
doi = {10.1088/0034-4885/74/7/076501},
url = {https://doi.org/10.1088/0034-4885/74/7/076501},
year = {2011},
month = {jun},
publisher = {},
volume = {74},
number = {7},
pages = {076501},
author = {Aharonovich, I and Castelletto, S and Simpson, D A and Su, C-H and Greentree, A D and Prawer, S},
title = {Diamond-based single-photon emitters},
journal = {Reports on Progress in Physics}
}

@article{BGN+2015,
	author = {Bradac, C. and Gaebel, T. and Naidoo, N. and Sellars, M. J. and Twamley, J. and Brown, L. J. and Barnard, A. S. and Plakhotnik, T. and Zvyagin, A. V. and Rabeau, J. R.},
	date = {2010/05/01},
	doi = {10.1038/nnano.2010.56},
	id = {Bradac2010},
	isbn = {1748-3395},
	journal = {Nature Nanotechnology},
	number = {5},
	pages = {345--349},
	title = {Observation and control of blinking nitrogen-vacancy centres in discrete nanodiamonds},
	url = {https://doi.org/10.1038/nnano.2010.56},
	volume = {5},
	year = {2010}}

@article{castelletto2020silicon,
  title = {Silicon Carbide Color Centers for Quantum Applications},
  author = {Castelletto, Stefania and Boretti, Alberto},
  year = {2020},
  journal = {Journal of Physics: Photonics},
  shortjournal = {J. Phys. Photonics},
  volume = {2},
  number = {2},
  pages = {022001},
  publisher = {IOP Publishing},
  issn = {2515-7647},
  doi = {10.1088/2515-7647/ab77a2},
  langid = {english}
}

@article{scholten2024multispecies,
  title = {Multi-Species Optically Addressable Spin Defects in a van Der {{Waals}} Material},
  author = {Scholten, Sam C. and Singh, Priya and Healey, Alexander J. and Robertson, Islay O. and Haim, Galya and Tan, Cheng and Broadway, David A. and Wang, Lan and Abe, Hiroshi and Ohshima, Takeshi and Kianinia, Mehran and Reineck, Philipp and Aharonovich, Igor and Tetienne, Jean-Philippe},
  year = {2024},
  journal = {Nature Communications},
  shortjournal = {Nat Commun},
  volume = {15},
  number = {1},
  pages = {6727},
  publisher = {Nature Publishing Group},
  issn = {2041-1723},
  doi = {10.1038/s41467-024-51129-8},
  langid = {english}
}

@article{sola-garcia2020electroninduced,
  title = {Electron-{{Induced State Conversion}} in {{Diamond NV Centers Measured}} with {{Pump}}--{{Probe Cathodoluminescence Spectroscopy}}},
  author = {Sol\`a-Garcia, Magdalena and Meuret, Sophie and Coenen, Toon and Polman, Albert},
  year = {2020},
  journal = {ACS Photonics},
  shortjournal = {ACS Photonics},
  volume = {7},
  number = {1},
  pages = {232--240},
  publisher = {American Chemical Society},
  doi = {10.1021/acsphotonics.9b01463}
}

@article{salehpour1990comparison,
  title = {Comparison of Electron Bands of Hexagonal and Cubic Diamond},
  author = {Salehpour, M. R. and Satpathy, S.},
  year = {1990},
  journal = {Physical Review B},
  shortjournal = {Phys. Rev. B},
  volume = {41},
  number = {5},
  pages = {3048--3052},
  publisher = {American Physical Society},
  doi = {10.1103/PhysRevB.41.3048}
}

@article{abdelghafar2026initio,
  title = {Ab Initio Study of {{N2V}} Color Center in Lonsdaleite/Diamond Dual Structure: {{A}} Two-Site {{Hubbard}} Model},
  shorttitle = {Ab Initio Study of {{N2V}} Color Center in Lonsdaleite/Diamond Dual Structure},
  author = {Abdelghafar, Khaled A. and Choi, Daniel S. and Askar, Khalid},
  year = {2026},
  journal = {Computational Materials Science},
  volume = {263},
  pages = {114412},
  issn = {0927-0256},
  doi = {10.1016/j.commatsci.2025.114412},
  urldate = {2025-12-04},
}

@article{CCG+2025,
	author = {Chen, Desi and Chen, Guwen and Lv, Long and Dong, Jiajun and Shang, Yuchen and Hou, Xuyuan and Wang, Yan and Shang, Jianqi and Wang, Saisai and Yin, Yankun and Liu, Ran and Zhang, Wei and Jiang, Zhou and He, Yan and He, Bingchen and Mao, Chengwen and Zhu, Shengcai and Sundqvist, Bertil and Liu, Bingbing and Yao, Mingguang},
	date = {2025/04/01},
	doi = {10.1038/s41563-025-02126-9},
	id = {Chen2025},
	isbn = {1476-4660},
	journal = {Nature Materials},
	number = {4},
	pages = {513--518},
	title = {General approach for synthesizing hexagonal diamond by heating post-graphite phases},
	url = {https://doi.org/10.1038/s41563-025-02126-9},
	volume = {24},
	year = {2025}}

\appendix

\end{document}


\title{Supplementary information \\
\titlestring}

\author{Giannis Thalassinos}
\affiliation{School of Science, RMIT University, Melbourne, VIC 3001, Australia}

\author{Alan G. Salek}
\affiliation{School of Science, RMIT University, Melbourne, VIC 3001, Australia}
\affiliation{RMIT Microscopy and Microanalysis Facility, RMIT University, Melbourne, VIC 3001, Australia}

\author{Daniel Stavrevski}
\affiliation{School of Science, RMIT University, Melbourne, VIC 3001, Australia}

\author{Qiang Sun}
\affiliation{School of Science, RMIT University, Melbourne, VIC 3001, Australia}

\author{Mitchell O. de Vries}
\affiliation{School of Science, RMIT University, Melbourne, VIC 3001, Australia}
\affiliation{Quantum Machines Unit, Okinawa Institute of Science and Technology, Onna, Okinawa 904-0495, Japan}

\author{Colin M. MacRae}
\affiliation{CSIRO Mineral Resources, Microbeam Laboratory, Clayton, VIC 3168, Australia}

\author{Nicholas C. Wilson}
\affiliation{CSIRO Mineral Resources, Microbeam Laboratory, Clayton, VIC 3168, Australia}

\author{Andrew G. Tomkins}
\affiliation{School of Earth, Atmosphere and Environment, Monash University, Melbourne, VIC 3800, Australia}

\author{Dougal G. Mcculloch}
\affiliation{School of Science, RMIT University, Melbourne, VIC 3001, Australia}
\affiliation{RMIT Microscopy and Microanalysis Facility, RMIT University, Melbourne, VIC 3001, Australia}

\author{Andrew D. Greentree}
\affiliation{School of Science, RMIT University, Melbourne, VIC 3001, Australia}

\maketitle

\renewcommand{\thefigure}{S\arabic{figure}}
\renewcommand{\thetable}{S\arabic{table}}

\section{Photoluminescence spectra}
\begin{figure}[tbh]
    \centering
    \includegraphics[width=0.95\linewidth]{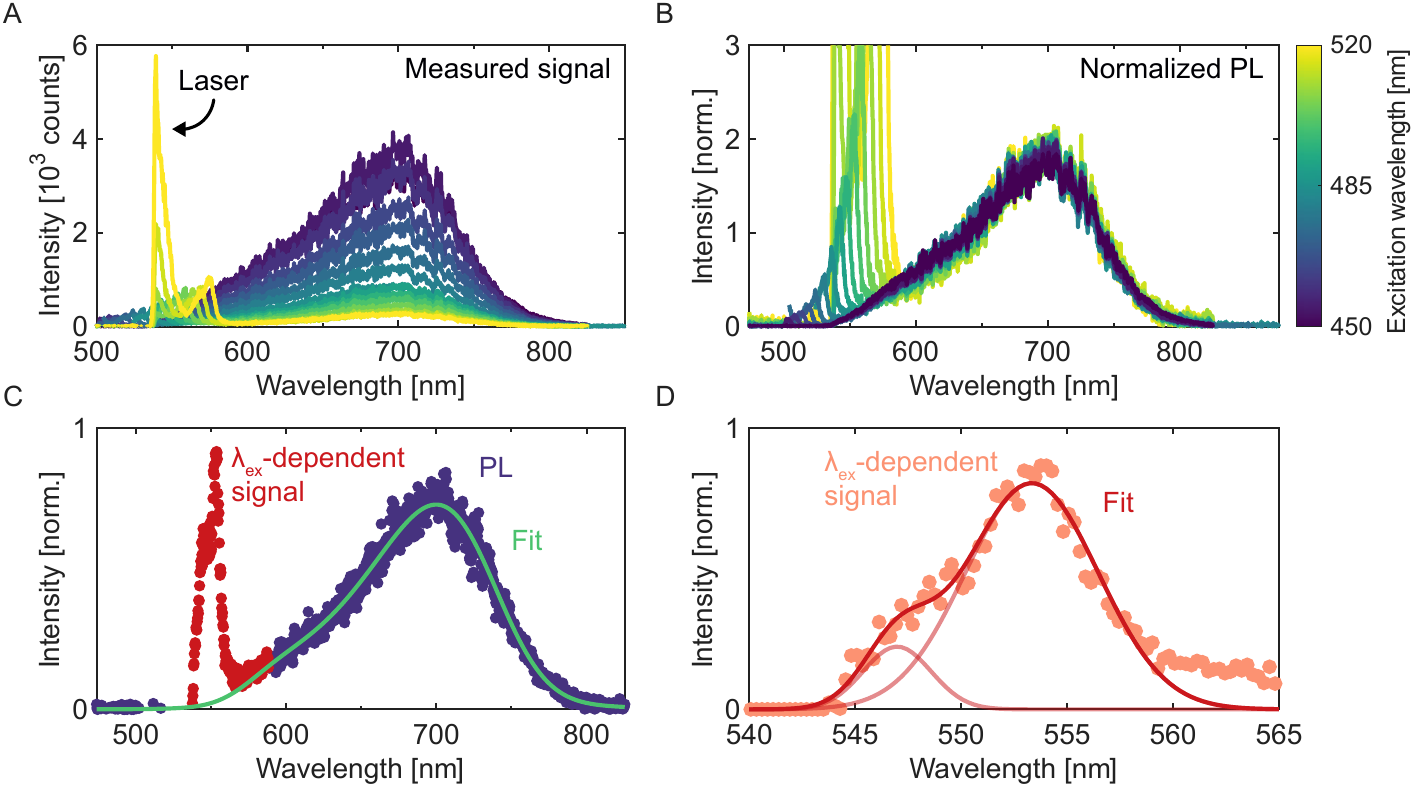}
    \caption{Photoluminescence of RU1.
    \textbf{A:}~As-measured photoluminescence of RU1. 
    The sharp, intense peak at longer excitation wavelengths corresponds to the laser bypassing the long-pass filter.
    \textbf{B:}~Photoluminescence of RU1 normalized between \numrange{600}{800}~nm, showing no change in spectral profile of the main signal. 
    \textbf{C:}~Example of double-skewed-Gaussian fit used to separate RU1 photoluminescence from the excitation wavelength-dependent feature.
    \textbf{D:}~Example of double-Gaussian fit used to calculate Stokes shifts. 
    }
    \label{fig:PL}
\end{figure}

The Lonsdaleite colour centre RU1 shows two overlapping signals, one of which changes only in intensity, which we attribute to RU1's photoluminescence (PL) spectrum, and an excitation wavelength (\lex)-dependent signal of undetermined origin. 
\Cref{fig:PL}A shows the as-measured photoluminescence spectra of RU1 before the signals were separated and after dark-counts subtraction. 
The bright, sharp feature which appears at longer excitation wavelengths corresponds to the laser beam bypassing the long pass filter. 
We show the same spectra normalized to the same photoluminescence intensity between \numrange{600}{800}~nm, which shows that RU1 exhibits no photochromism (\cref{fig:PL}B). 

To separate the two signals, we performed a double-skewed Gaussian fit of each RU1 PL spectrum for each \lex, ignoring the \lex-dependent signal.
An example can be seen in \cref{fig:PL}C. 
We then performed a double-Gaussian fit on the \lex-dependent signal to determine the Stokes shift of the two peaks, as shown in \cref{fig:PL}D.

\section{Elemental maps of region 2}
\begin{figure}[tbh]
    \centering
    \includegraphics[width=1\linewidth]{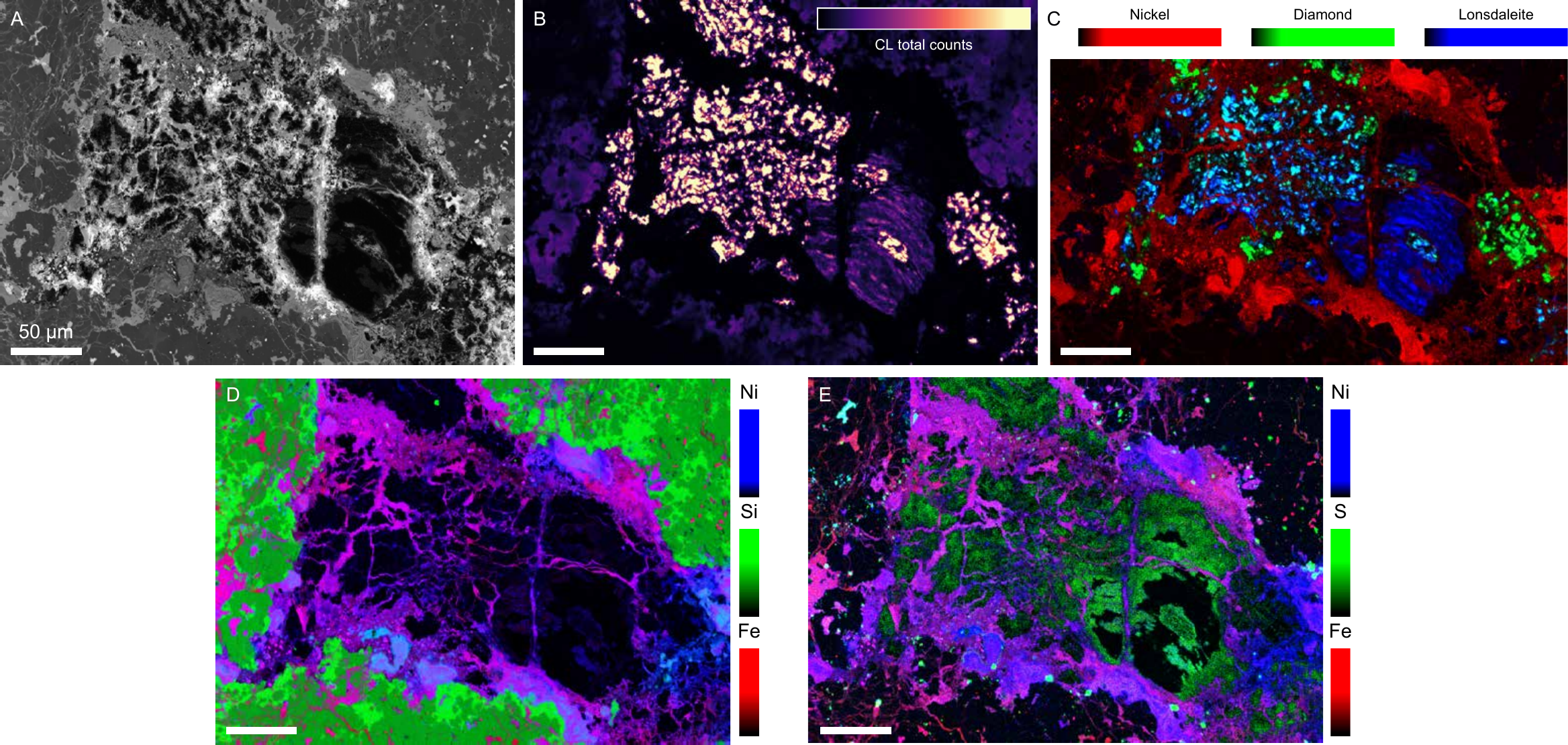}
    \caption{
    Maps corresponding to the second lonsdaleite region.
    \textbf{A:}~Scanning electron microscopy map.
    \textbf{B:}~Full CL map shown in figure 3A of the main text using total CL intensity.
    \textbf{C:}~CL map showing nickel, diamond, and lonsdaleite.
    \textbf{D:}~Elemental map showing the presense of Ni, Fe, and Si or S (E).
    The scale bar is the same across all panels.
    } 
    \label{fig:cl_maps}
\end{figure}

In figure 3 of the main text, we present additional data corresponding to a second Lonsdaleite grain. 
Here, we show additional maps corresponding to this same region of interest. 
\Cref{fig:cl_maps}A shows the region of interest under a scanning electron microscope (SEM) with \cref{fig:cl_maps}B showing the cathodoluminescence (CL) map by total counts. 
\Cref{fig:cl_maps}C is another CL map, but instead of total counts, we used the features identified in the CL spectra shown in fig. 1E of the main text to create a composite image, further distinguishing diamond, Lonsdaleite, and the surrounding Ni. 
Similarly, we show two elemental maps (\cref{fig:cl_maps}D \& E) which show the presence of Ni, Si, Fe, and S in the same region of interest as shown in fig. 3 of the main text.

